\documentclass[preprint, aip, pof, superscriptaddress]{revtex4-1}

\usepackage{times}
\usepackage{amssymb,amsmath}
\usepackage{subfigure}

	\usepackage[usenames,dvipsnames]{color}
	\usepackage[pdftex]{graphicx}
	\usepackage{epstopdf}
	\usepackage[pdftex]{hyperref}
	\hypersetup{
    	pdftitle={Manuscript on thin nematic film: strong anchoring model}, 
    	pdfauthor={TL, LJC, AA, LK and UT},  %ADAPT TEXT 
	pdfproducer={LaTeX},
	pdfview=FitV,       % FitH
	pdfstartview=FitB,
	linkcolor=blue,     % links to same page
	citecolor=blue,     % citations
	urlcolor=red,      % links to URLs
	breaklinks=true,    % links may be split onto 2 lines
	colorlinks=true,
	citebordercolor=0 0 0,  % color for \cite
	filebordercolor=0 0 0,
	linkbordercolor=0 0 0,
	menubordercolor=0 0 0,
	urlbordercolor=0 0 0,
	pdfhighlight=/I,
	pdfborder=0 0 0,   % no box around links
	bookmarksopen=true,
	bookmarksnumbered=true
	}
	\DeclareGraphicsExtensions{.pdf, .eps}

\newcommand{\vect}[1]{\mbox{\boldmath $#1$}}
\newcommand{\mylab}[1]{\label{#1}}
\begin{document}
%%%%%%%%%%%%%%%%%%%%%%%%%%%%%%%%%%%%%%%%%
%  Title and authors
%%%%%%%%%%%%%%%%%%%%%%%%%%%%%%%%%%%%%%%%%
\title{Note on the hydrodynamic description of thin nematic films: strong anchoring 
model}

\author{Te-Sheng Lin}
\email{t.lin@lboro.ac.uk}
\affiliation{Department of Mathematical Sciences, Loughborough University, 
Loughborough, Leicestershire LE11 3TU, UK}
\author{Linda J. Cummings}
\email{linda.cummings@njit.edu}
\affiliation{Department of Mathematical Sciences and Center for Applied 
Mathematics and Statistics, New Jersey Institute of Technology, Newark, New 
Jersey 07102, USA}
\author{Andrew J. Archer}
\email{a.j.archer@lboro.ac.uk}
\affiliation{Department of Mathematical Sciences, Loughborough University, 
Loughborough, Leicestershire LE11 3TU, UK}
\author{Lou Kondic}
\email{kondic@njit.edu}
%\homepage{http://m.njit.edu/~kondic/}
\affiliation{Department of Mathematical Sciences and Center for Applied 
Mathematics and Statistics, New Jersey Institute of Technology, Newark, New 
Jersey 07102, USA}
\author{Uwe Thiele}
\email{u.thiele@lboro.ac.uk}
%\homepage{http://www.uwethiele.de}
\affiliation{Department of Mathematical Sciences, Loughborough University, 
Loughborough, Leicestershire LE11 3TU, UK}

\date{\today}

%%%%%%%%%%%%%%%%%%%%%%%%%%%%%%%%%%%%%%%%%
%  Abstract
%%%%%%%%%%%%%%%%%%%%%%%%%%%%%%%%%%%%%%%%%

\begin{abstract}
We discuss the long-wave hydrodynamic model for a thin film of nematic liquid 
crystal in the limit of strong anchoring at the free surface and at the substrate. 
We rigorously clarify how the elastic energy enters the evolution equation for 
the film thickness in order to provide a solid basis for further investigation: 
several conflicting models exist in the literature that predict qualitatively 
different behaviour. We consolidate the various approaches and show that the 
long-wave model derived through an asymptotic expansion of the full 
nemato-hydrodynamic equations with \textit{consistent} boundary conditions 
agrees with the model one obtains by employing a thermodynamically motivated 
gradient dynamics formulation based on an underlying free energy functional. As 
a result, we find that in the case of strong anchoring the elastic distortion energy 
is always stabilising. To support the discussion in the main part of the paper, an 
appendix gives the full derivation of the evolution equation for the film thickness 
via asymptotic expansion.
\end{abstract}

\maketitle

%%%%%%%%%%%%%%%%%%%%%%%%%%%%%%%%%%%%%%%%%
\section{Introduction} \mylab{sec:INTRO}
%%%%%%%%%%%%%%%%%%%%%%%%%%%%%%%%%%%%%%%%%

Thin films of nematic liquid crystals (NLC) have attracted attention over the years, 
as evidenced by a number of experimental and theoretical 
studies~\cite{Lavrentovich1994, Sparavigna1994, Lavrentovich1995, Delabre2008, 
Delabre2010, Cazabat2011, BenAmar2001, Effenterre2003, Mechkov2009a, 
Manyuhina2010}. When thin nematic films are deposited on solid or liquid 
substrates, they often exhibit antagonistic anchoring at the free surface and at the 
substrate, i.e., the director orientation at the substrate is generally parallel to the 
substrate (planar anchoring) but at the free surface the director is orthogonal to the 
surface (homeotropic anchoring). As a consequence, the local director orientation 
changes across the film resulting in an elastic contribution to the energy that 
should not be neglected: such films are called hybrid films. Sometimes instabilities 
are observed that result in lateral periodic stripe patterns of the director 
orientation~\cite{Lavrentovich1994, Sparavigna1994, Lavrentovich1995, 
Delabre2008, Delabre2010, Cazabat2011} and film height. However, this is only the 
case for thin films with thicknesses of several hundred nanometer and below; the 
wavelength of the stripe patterns diverges at an upper critical film thickness and so, 
for thicker films, only the usual defects of the nematic phases are 
observed~\cite{Lavrentovich1995, Delabre2010}. Note that spinodal patterns have 
also been observed~\cite{Effenterre2003, Herminghaus1998, Schlagowski2002, 
Garcia2008}, normally, in the vicinity of nematic-isotropic or smectic-nematic phase 
transitions. In contrast, the stripe patterns are observed well inside the nematic 
region of the liquid crystal phase diagram.

In order to develop a theory for the behaviour of confined nematic liquid crystals, 
one may calculate the director orientation profile for a given static free surface. 
Typically, either a flat film or a periodically deformed state is considered. Such a 
given static geometry is then used to investigate the director field and to determine 
its stability. For an imposed flat film, an energy argument allows one to show that 
there exists a critical thickness~\cite{Barbero1983}
\begin{equation}
h_c= \left |\frac{K}{A_+}-\frac{K}{A_-} \right|,
\mylab{eq:critical}
\end{equation}
where $K$ is the bulk elastic constant of the liquid crystal (in the one constant 
approximation) and $A_+$ and $A_-$ are the anchoring strengths at the free 
surface and at the substrate, respectively. For thin films, with thickness $h\le h_c$, 
the director profile is undistorted; the film is in the so-called planar (P) state and 
the director is aligned parallel to the anchoring angle at the interface with the 
stronger anchoring strength. For thick films, with film thickness $h>h_c$, the state 
that minimises the free energy is that where the director orientation changes 
continuously between the two anchoring directions as one moves across the film; 
this is the Hybrid-Aligned-Nematic (HAN) state introduced above and is the case 
for the strong anchoring situation considered here. If one assumes that the system is 
invariant in one direction across the surface on which the film is deposited, so that 
the film is effectively two-dimensional (2D), one finds that these states are linearly 
stable. To confirm this assumption, much effort has gone into determining whether 
the film is laterally stable~\cite{Lavrentovich1994, Sparavigna1994, Lavrentovich1995, 
Manyuhina2010}. However, since the film geometry is imposed and static, such 
analyses can not account for a possible coupling of variations in film height and 
director orientation.

In alternative approaches, the long-wave hydrodynamic or so-called lubrication 
theory has been used successfully in deriving the film thickness evolution equations 
for films of a variety of different (simple) liquids and to explore the dynamics 
under the influence of gravity or other body forces, and a variety of surface 
and interfacial forces~\cite{Oron1997, Ruschak1999, Chang2002, Kondic2003, 
Stone2004, Craster2009, Thiele2010}. In order to extend this approach to describe 
films of NLCs,  
Ben Amar and Cummings~\cite{BenAmar2001} derived a model to describe the 
surface evolution of NLCs with strong anchoring in 2D settings that was later 
adapted to model 2D spreading droplets~\cite{Cummings2011}, spreading 
droplets with defects~\cite{Lin2012} and to account for three dimensional settings 
(3D)~\cite{Cummings2004}.  Another long-wave model was introduced by Carou 
et al.\ to study blade coating and cavity filling flows of NLC in 2D~\cite{Carou2005, 
Carou2006, Carou2007}. However, none of these long-wave evolution equations 
agree with models that use energy arguments~\cite{Mechkov2009a}, 
when it comes to 
identifying the effect of the elastic distortion energy on the film dynamics. 
Antagonistic anchoring is predicted to destabilise the film in 
Refs.~\onlinecite{BenAmar2001, Cummings2004, Cummings2011, Lin2012}, but in 
Refs.~\onlinecite{Carou2005, Carou2006, Carou2007} it is predicted to have no 
influence on the stability of the film. 
In Ref.~\onlinecite{Mechkov2009a,Vandenbrouck1999}, however, it is argued on 
physical grounds that the elastic energy is stabilising. Thus, predictions based on 
the theory of 
Refs.~\onlinecite{BenAmar2001, Cummings2004, Cummings2011, Lin2012} are in 
direct conflict with those from the theory in 
Refs.~\onlinecite{Mechkov2009a} and~\onlinecite{Vandenbrouck1999}. 

On a different note, the energetic approach to deriving the long-wave theory 
mentioned above is based on the fact that, as was noted some time ago, the 
evolution equation for the height of a thin Newtonian film can be written in a 
variational form in situations where inertia can be neglected. For nematic liquid 
crystals, it is not a priori clear whether or not this approach can be applied. In 
Ref.~\onlinecite{Mechkov2009a}, a model is derived based on an energy 
argument and a gradient dynamics ansatz that employs a mobility typically for 
isotropic liquids. However, no mathematical justification was given. 

The purpose of this note is to clarify these issues by reconciling 
the hydrodynamic long-wave and energetic approaches in the case of layers of 
nematic liquid crystals with strong anchoring, and so to provide a solid basis for 
further investigations. Our main results are as follows: (1) In the case of strong 
antagonistic anchoring, the elastic energy contribution always acts so as to 
stabilise the layer. This is found employing the long-wave approximation of the 
governing nematohydrodynamic bulk equations with consistent interfacial 
boundary conditions, and as well by employing a thermodynamically motivated 
gradient dynamics formulation. (2) The long-wave models of 
Refs.~\onlinecite{BenAmar2001, Cummings2004, Cummings2011, Lin2012} 
were derived by employing a stress balance at the free surface of the film as in 
standard Newtonian flow, that is inconsistent with the bulk equations. This leads 
to a change in sign of the elastic contribution in the film thickness evolution 
equation. When this boundary condition is modified to also include the elastic 
stress, results consistent with the energy approach are obtained. 
(3) The mobility function in the gradient dynamics approach must be obtained 
from hydrodynamics. Here we show that the evolution equation for the height 
of a thin film of nematic liquid crystals derived via asymptotic expansion of the 
full nematohydrodynamic equations can be written in a variational form and so 
is consistent with the gradient dynamics approach. 

The manuscript is organised as follows: The continuum theory of NLC,
including the elastic energy and Ericksen-Leslie bulk equations together
with consistent boundary conditions, is given in
Section~\ref{sec:CONTI}. Focusing on the 2D case, the long-wave
approximation of the governing equations and boundary conditions is
sketched in Section~\ref{sec:longwave} while the full details are
given in Appendix~\ref{app:longwave}. This allows the reader to easily
reproduce our main findings. In Section~\ref{sec:GDF}, a
thermodynamically motivated gradient dynamics formulation is employed
to derive the evolution equation of a nematic film. The stability of
the free surface is studied through a linear stability
analysis. Finally, in Section~\ref{sec:DC} we compare the results of
the two approaches and discuss the validity and limitation of the
present model. The note concludes with an outlook on related problems
that could be studied based on our results.

%%%%%%%%%%%%%%%%%%%%%%%%%%%%%%%%%%%%%%%%%
\section{Continuum description of nematic liquid crystal} \mylab{sec:CONTI}
%%%%%%%%%%%%%%%%%%%%%%%%%%%%%%%%%%%%%%%%%

Nematic liquid crystals consist of rod-like molecules that have no positional 
order, but have long-range orientational order. Thus, the molecules are free to flow 
as a liquid, but still maintain their long-range directional order. The mean molecule 
alignment is described by the unit vector $\vect{n}=(n_1,n_2,n_3)^T$ where the 
superscript $T$ denotes matrix transposition. Further notation conventions used 
here are presented in Appendix~\ref{sec:NOTA}.

Distortions of the director field result in a contribution to the free energy, 
that for NLC is known as the Frank-Oseen elastic energy and 
reads~\cite{DeGennes1995, Stewart2004}
\begin{eqnarray}
w_F &=& \frac{1}{2} 
K_1(\nabla\cdot \vect{n})^2 
+ \frac{1}{2} K_2 (\vect{n}\cdot\nabla\times\vect{n})^2 
+ \frac{1}{2} K_3 (\vect{n}\times\nabla\times\vect{n})^2 \nonumber\\
&& + \frac{1}{2} (K_2+K_4) \nabla\cdot\left((\vect{n}\cdot\nabla)\vect{n}
-(\nabla\cdot\vect{n})\vect{n}\right),
\end{eqnarray}
where $K_1$, $K_2$ and $K_3$ are the splay, twist and bend elastic constants, 
respectively, and $(K_2+K_4)$ is called the saddle-splay constant. Note that the 
saddle-splay term is often omitted since it does not contribute to the governing 
equations in the case of strong anchoring.

We use the one-constant approximation to simplify the problem. One 
assumes~\cite{DeGennes1995, Stewart2004}
\begin{equation}
K\equiv K_1= K_2= K_3, \quad K_4=0.
\end{equation}
and obtains the simplified energy density
\begin{equation}
w_F =\frac{K}{2}\, \nabla\vect{n} : (\nabla\vect{n})^T 
= \frac{K}{2}\, n_{l,k}n_{l,k} 
\end{equation}
that enters the nemato-hydrodynamic equations discussed next.

\subsection{Ericksen-Leslie equation} \mylab{sec:ELF}

The bulk flow of NLC may be described by the Ericksen-Leslie 
equations~\cite{Ericksen1959, Leslie1968,DeGennes1995, Stewart2004}. The 
fluid is incompressible, satisfying 
\begin{equation}
\nabla\cdot \vect{v}=0,
\mylab{eq:ELFI}
\end{equation}
where $\vect{v}=(v_1,v_2,v_3)^T$ is the velocity field. The momentum balance 
equation is
\begin{equation}
\rho \frac{D}{Dt}\vect{v} = \nabla\cdot\vect{\sigma},
\mylab{eq:ELFL}
\end{equation}
where $\rho$ is the density, $D/Dt=\partial/\partial t + \vect{v}\cdot\nabla$ is the 
material derivative, $t$ is the time variable and $\vect{\sigma}$ is the stress 
tensor of the NLC. The stress tensor is defined as~\cite{DeGennes1995}
\begin{equation}
\vect{\sigma} = -p\,\vect{I} + \vect{\sigma}^E + \vect{\sigma}^V,
\mylab{eq:ELFS1}
\end{equation}
where $p$ is the pressure, $\vect{I}$ is the identity tensor, $\vect{\sigma}^E$ is 
the elastic (Ericksen) stress tensor, defined by
\begin{equation}
\vect{\sigma}^E =  
-\frac{\partial w_F}{\partial \nabla\vect{n}} \cdot (\nabla\vect{n})^T,
\end{equation}
and $\vect{\sigma}^V$ is the viscous stress tensor with components
\begin{equation}
\sigma^V_{ij}=\alpha_1 n_kn_p e_{kp}n_in_j + \alpha_2 N_in_j+
\alpha_3 N_jn_i+\alpha_4 e_{ij}+\alpha_5 e_{ik}n_kn_j+\alpha_6 e_{jk}n_kn_i,
\end{equation}
where
\begin{equation}
e_{ij}=\frac{1}{2}\left(v_{i,j}+v_{j,i}\right), \quad
w_{ij}=\frac{1}{2}\left(v_{i,j}-v_{j,i}\right), \quad 
N_{i}=\frac{D}{Dt}n_i-w_{ik}n_k.
\end{equation}
The $\alpha_i$ are constant viscosities. 

The equation for the balance of angular momentum is written as (neglecting 
director inertia)
\begin{equation}
\nabla\cdot\left(\frac{\partial w_F}{\partial \nabla\vect{n}}\right)
-\frac{\partial w_F}{\partial \vect{n}}+\vect{g}=\lambda\vect{n}, 
\mylab{eq:ELFA}
\end{equation}
where the components of $\vect{g}$ are
\begin{equation}
g_i=-\gamma_1 N_i - \gamma_2 e_{i,k} n_k, \quad 
\gamma_1=\alpha_3-\alpha_2, \quad \gamma_2=\alpha_6-\alpha_5.
\end{equation}
Furthermore, $\lambda$ is the Lagrange multiplier ensuring $|\vect{n}|=1$.

Under the assumption of the one constant approximation, the Ericksen-Leslie 
equations, Eqs.~(\ref{eq:ELFI}), (\ref{eq:ELFL}), and (\ref{eq:ELFA}) simplify to
\begin{subequations}
\begin{eqnarray}
\nabla\cdot\vect{v} &=& 0, \mylab{eq:ELKI} \\
\rho \frac{D}{Dt}\vect{v} &=& 
-\nabla(p+w_F) -K\, \nabla\vect{n}\cdot\Delta\vect{n}+\nabla\cdot\vect{\sigma}^V, 
\mylab{eq:ELKL} \\
K \Delta \vect{n}+\vect{g} &=& \lambda \vect{n}, \mylab{eq:ELKA}
\end{eqnarray}
\mylab{eq:ELK}
\end{subequations}
respectively, where we have used that
\begin{equation}
\vect{\sigma}^E = -K\,\nabla\vect{n}\cdot (\nabla\vect{n})^T,
\end{equation}
and
\begin{equation}
\nabla\cdot(\nabla\vect{n} \cdot (\nabla\vect{n})^T)=
\frac{1}{2}\nabla\left(\nabla\vect{n} : (\nabla\vect{n})^T\right)
+\nabla\vect{n}\cdot\Delta\vect{n}.
\end{equation}
As a result, the Ericksen-Leslie equations in the one constant approximation 
are given by Eq.~(\ref{eq:ELK}) and need to be solved subject to appropriate 
boundary conditions.

\paragraph{Remark 1:}
Note that sometimes the stress tensor for NLC is written differently from 
Eq.~(\ref{eq:ELFS1}), e.g., Ref.~\onlinecite{Rey2000} uses
\[
\tilde{\vect{\sigma}} = -(\tilde p+w_F)\,\vect{I} + \vect{\sigma}^E + \vect{\sigma}^V.
\]
However, one may combine the two terms of the isotropic part of 
$\tilde{\vect{\sigma}}$ and define a modified pressure as $p=\tilde{p}+w_F$. 
Hence, with the exception of the modified pressure the derivations that follow 
are not affected.

\paragraph{Remark 2:}
Equation~(\ref{eq:ELKL}) can be rewritten as
\[
\rho \frac{D}{Dt}\vect{v} = -\nabla(p+w_F) 
+\nabla\vect{n}\cdot\vect{g}+\nabla\cdot\vect{\sigma}^V
\]
by using Eq.~(\ref{eq:ELKA}) together with $\nabla\vect{n}\cdot\vect{n}=\vect{0}$. 
This formulation is more popular in the literature since it only involves the first 
derivative of the director field.

\subsubsection{Boundary conditions}

We assume here that the NLC film sits on a solid substrate at $z=0$ with the 
free surface (or film thickness) described by $z=h(x,y,t)$.

For the director field $\vect{n}$, we impose strong anchoring conditions such 
that the director is planar at the solid substrate and is homeotropic at the 
free surface. Specifically, we have
\begin{subequations}
\begin{eqnarray}
& \vect{n}\cdot \vect{z}=0, & \quad \mbox{at $z=0$}, \mylab{eq:ELKB1}\\
& \vect{n}\cdot \vect{t_i}=0, & \quad \mbox{at $z=h(x,y,t)$}, \mylab{eq:ELKB2}
\end{eqnarray}
\end{subequations}
where $\vect{z}=(0,0,1)^T$ and $\vect{t_i}$ are the surface tangent vectors, 
\begin{equation}
\vect{t_1}=\frac{1}{\sqrt{1+(\partial_x h)^2+(\partial_y h)^2}}(1,0,\partial_x h)^T, \quad 
\vect{t_2}=\frac{1}{\sqrt{1+(\partial_x h)^2+(\partial_y h)^2}}(0,1,\partial_y h)^T.
\end{equation}

For the velocity field $\vect{v}$, we assume no-slip and no-penetration at the solid substrate,
\begin{equation}
v_1=v_2=v_3=0, \quad \mbox{at $z=0$}.
\end{equation}
At the free surface, $z=h(x,y,t)$, we have the kinematic condition and balance 
of normal and tangential stresses. The kinematic condition is 
\begin{equation}
v_3 = \partial_t h + v_1\partial_x h + v_2 \partial_y h, \quad 
\mbox{at $z=h$.}
\end{equation}
For normal stress, we assume that the jump across the interface is balanced 
by surface tension. That is,
\begin{equation}
\vect{k}\cdot(\vect{\sigma}-\vect{\sigma}^i)\cdot \vect{k}=2\gamma H, \quad 
\mbox{at $z=h$,}
\end{equation}
where  
$\vect{\sigma}^i=-p_0\vect{I}$ is the stress tensor of the air phase, $p_0$ is the 
atmospheric pressure, $\gamma$ is the surface tension, $H$ is the mean 
curvature and $\vect{k}$ is the surface normal vector
\begin{equation}
\vect{k} = \frac{1}{\sqrt{1+(\partial_x h)^2+(\partial_y h)^2}}\,
\left(-\partial_xh,-\partial_yh,1\right)^T.
\end{equation}
For tangential stress, we assume that there is no jump at the interface 
\begin{equation}
\vect{k}\cdot (\vect{\sigma}-\vect{\sigma}^i) \cdot \vect{t_i}=0.
\end{equation}
That is, we assume that no tangential surface tension gradient exists, as is 
appropriate for strong anchoring. For the case where surface gradient exists, 
see Ref.~\onlinecite{Rey1999}.

%%%%%%%%%%%%%%%%%%%%%%%%%%%%%%%%%%%%%%%%%
\section{Long-wave hydrodynamic description in two dimensions} 
\mylab{sec:longwave}
%%%%%%%%%%%%%%%%%%%%%%%%%%%%%%%%%%%%%%%%%

In this section we restrict attention to two space dimensions and focus on the 
long-wave approximation of the governing equations presented previously in 
Sec.~\ref{sec:CONTI}. The full details are given in Appendix~\ref{app:longwave}. 
The aim here is to study the contribution of nematic elasticity to the free surface 
evolution, and to distinguish results obtained using different scalings and boundary 
conditions.

Assume the flow is two dimensional and $y$-independent, so that the director field 
can be expressed as $\vect{n}=(\sin\theta,\cos\theta)^T$ where the angle $\theta$ 
is taken as the difference between the director orientation and the positive $z$-axis, 
as shown in Fig.~\ref{FIG:director}(a), and the velocity field is $\vect{v}=(u,w)^T$. 
We introduce long-wave scalings to nondimensionalize the governing 
equations. The scalings are
\begin{equation}
(x,z)=(L\bar{x}, \delta L\bar{z}), \quad 
(u,w)=(U\bar{u},\delta U\bar{w}), \quad 
t=\frac{L}{U}\,\bar{t}, \quad 
p=\frac{\mu U}{\delta^2 L}\,\bar{p},
\mylab{eq:2DSCAL}
\end{equation}
where $U$ is the scale of fluid velocity, $\delta=H/L \ll 1$ is the ratio between the 
typical film thickness scale, $H$, and a typical lateral length scale, $L$. In addition, 
in order to focus only on the nematic elasticity, we approximate the nematic viscous 
stress tensor by its Newtonian equivalent, setting $\sigma^V_{ij} = 2\mu e_{ij}$, 
where $\mu=\alpha_4/2$.

\subsection{Weak elasticity} \mylab{sec:2DS1}

Assuming that the elastic free energy is weak compared to the pressure, we can 
introduce the dimensionless number (inverse Ericksen number)
\begin{equation}
\bar{K} = \frac{K}{\delta\mu U L}.
\end{equation}
The leading order bulk equations are then given by (after dropping the over-bars)
\begin{subequations}
\begin{eqnarray}
\partial_x p &=& \partial^2_{z} u, \\
\partial_zp &=& 0, \\
K\,\partial^2_z\theta &=& 0, \\
\partial_xu + \partial_zw &=& 0.
\end{eqnarray}
\end{subequations}
In addition, the leading order boundary conditions are 
\begin{equation}
\theta(z=0)=\frac{\pi}{2}, \quad \theta(z=h)=0, 
\end{equation}
\begin{equation}
p = p_0-C\,\partial^2_xh, \quad \partial_zu=0, \quad \mbox{at $z=h$},
\mylab{eq:2DS1B1}
\end{equation}
where $C=\delta^3 \gamma/\mu U$ is the inverse capillary number.

It is easily seen that the velocity field and the director field are decoupled. 
The film thickness evolution equation is obtained as
\begin{equation}
\partial_th + \partial_x\left( \frac{C}{3}\,h^3 \,\partial^3_xh\right) =0,
\mylab{eq:TFCDMW}
\end{equation}
and the director field satisfies
\begin{equation}
\theta = \frac{\pi}{2}\left(1-\frac{z}{h}\right).
\mylab{eq:director}
\end{equation}
This corresponds to the approach taken in Refs.~\onlinecite{Carou2005, Carou2006, 
Carou2007}.

\subsection{Moderate elasticity} \mylab{sec:2DS2}

Instead, if we introduce the inverse Ericksen number as
\begin{equation}
\bar{K} = \frac{K}{\mu U L},
\end{equation}
the leading order bulk equations are given by (after dropping the over-bars)
\begin{subequations}
\begin{eqnarray}
\partial_x \left(p+\frac{K}{2}\,(\partial_z\theta)^2\right) &=& \partial^2_zu, \\
 \partial_zp &=& 0, \\
 K\,\partial^2_z\theta &=& 0, \\
 \partial_xu + \partial_zw &=& 0,
\end{eqnarray}
\mylab{eq:2DS2EL}
\end{subequations}
with leading order boundary conditions
\begin{equation}
\theta(z=0)=\frac{\pi}{2}, \quad \theta(z=h)=0,
\mylab{eq:2DS2B1}
\end{equation}
\begin{equation}
p = p_0-C\,\partial^2_xh-K\,(\partial_z\theta)^2, \quad
-K\,(\partial_x\theta\,\partial_z\theta + (\partial_z\theta)^2\,\partial_xh)+\partial_zu =0, 
\quad \mbox{at $z=h$}.
\mylab{eq:2DS2B2}
\end{equation}

Under such a scaling, the director field is decoupled from the flow and is given by 
Eq.~(\ref{eq:director}). The tangential stress boundary condition in 
Eq.~(\ref{eq:2DS2B2}) is then reduced to $\partial_zu(z=h)=0$. We can therefore solve 
for the pressure and velocity field exactly. As a result, the film evolution equation 
is given by 
\begin{equation}
\partial_th + \partial_x\left(\frac{C}{3}\,h^3 \,\partial^3_xh 
-\frac{\widetilde{K}}{3}\,\partial_xh\right) =0,
\mylab{eq:TF2D}
\end{equation}
where $\widetilde{K}=\pi^2 K/4$.

In contrast, Ben Amar \& Cummings~\cite{BenAmar2001} employ 
Eqs.~(\ref{eq:2DS2EL}, \ref{eq:2DS2B1}) and impose the normal stress balance 
assuming that the jump of the pressure is balanced by surface tension alone, as is 
appropriate for Newtonian fluids; that 
is, they use Eq.~(\ref{eq:2DS1B1}) instead of Eq.~(\ref{eq:2DS2B2}). As a result, 
they obtain an equation much like Eq.~(\ref{eq:TF2D}) but with the opposite sign 
for the elasticity term. This issue will be discussed later in Sec.~\ref{sec:DC}.

%%%%%%%%%%%%%%%%%%%%%%%%%%%%%%%%%%%%%%%%%
\section{Gradient dynamics formulation for a thin film of nematic liquid crystals in 
two dimensions} \mylab{sec:GDF}
%%%%%%%%%%%%%%%%%%%%%%%%%%%%%%%%%%%%%%%%%

It was noted some time ago that the time evolution equation for the height of a 
thin Newtonian film on a solid substrate can be written in a variational form in 
situations where inertia can be neglected~\cite{Oron1992, S.M.1993, Thiele2010}. 
The evolution of the film thickness $h$ follows a dissipative gradient dynamics 
governed by equation 
\begin{equation}
\partial_th = \partial_x\left[Q(h)\,\partial_x\left(\frac{\delta F}{\delta h}\right)\right],
\mylab{eq:GDF2D}
\end{equation}
where $\delta/\delta h$ denotes functional variation with respect to $h$. The 
resulting relaxation dynamics is governed by the free energy functional $F$ with the 
mobility function $Q(h)$.

Such an approach may also be used to obtain the evolution equation for a NLC 
film in the limit of moderate elasticity discussed above in 
section~\ref{sec:2DS2}. Restricting our attention again to a 2D geometry, we 
simplify the elastic distortion energy for the case of lateral long-wave distortions, 
i.e., we assume the scalings given in Eq.~(\ref{eq:2DSCAL}). The bulk elastic 
energy is to leading order
\begin{equation}
w_F =\frac{K}{2}\,(\partial_z\theta)^2. 
\end{equation}
We further assume that the director adjusts instantaneously to its steady state as 
compared to the fluid relaxation time, i.e., we assume $K=O(\mu U L)$ . Then, 
the director field can be exactly solved for to obtain a linear profile as shown in 
Eq.~(\ref{eq:director}) assuming strong planar anchoring at the solid substrate 
and strong homeotropic anchoring at the free surface. The corresponding director 
orientation across the film is sketched in Fig.~\ref{FIG:director}(b).

As a result, the bulk elastic energy $w_F$ (energy/volume) of the NLC can be 
rewritten as:
\begin{equation}
w_F=\frac{\widetilde{K}}{2h^2}.
\end{equation}
The free energy functional is then expressed as
\begin{equation}
F = \int C\,ds + \int \left(\int^h_0 w_F \, dz \right)\, dx \approx
\int \left[C \left(1+\frac{(\partial_x h)^2}{2}\right) +\frac{\widetilde{K}}{2h}\right]\,dx,
\mylab{eq:GDFSA}
\end{equation}
where $ds\approx (1+(\partial_xh)^2/2) \, dx$ is the 
approximated surface element. The evolution equation of the film is given in 
gradient dynamics formulation by introducing $F$ into Eq.~(\ref{eq:GDF2D}):
\begin{equation}
h_t = -\partial_x\left[Q(h)\,\partial_x\left(C\, \partial^2_xh +
\frac{\widetilde{K}}{2 h^2}\right)\right],
\mylab{eq:GDTF}
\end{equation}
where the mobility function $Q(h)$ can be 
obtained from the Poiseuille NLC flow, Eq.~(\ref{eq:S2TF2}). One should note 
that Eq.~(\ref{eq:GDTF}) and Eq.~(\ref{eq:TF2D}) are identical when $Q(h)=h^3/3$.

\subsection{Linear stability analysis}

To have a basic understanding of the elastic contribution to the stability of NLC 
free surface, we analyse the linear stability of a flat film, $h=h_0$. Assuming 
$h=h_0+\xi$, $\xi\ll h_0$ in Eq.~(\ref{eq:GDTF}), to leading order we have
\begin{equation}
\partial_t\xi = -Q(h_0) \left(C\, \partial^4_x\xi - \frac{\widetilde{K}}{h^3_0}\,
\partial^2_x\xi\right).
\end{equation}
With the harmonic mode ansatz $\xi=\exp(ikx+\omega t)$ one obtains the 
dispersion relation
\begin{equation}
\omega = -Q(h_0)\left(C\, k^4 + \frac{\widetilde{K}}{h^3_0}\,k^2\right).
\end{equation}
Note that the constants $C$, $\tilde K$ and the film height $h_0$ are always 
positive and therefore the growth rate $\omega$ is negative for any 
wavenumber $k$. This implies that the elastic term is always stabilising and in 
the case of strong anchoring the flat film $h=h_0$ is always stable if only 
capillarity and elasticity are taken into account.

%%%%%%%%%%%%%%%%%%%%%%%%%%%%%%%%%%%%%%%%%
\section{Discussion and conclusion} \mylab{sec:DC}
%%%%%%%%%%%%%%%%%%%%%%%%%%%%%%%%%%%%%%%%%

We have consolidated several approaches to derive the evolution equation for free 
surface films of nematic liquid crystals with strong anchoring at both interfaces. We 
have demonstrated the consistency between the long-wave approximation model, 
Eq.~(\ref{eq:TF2D}) and Eq.~(\ref{eq:TFNLC}), and the model derived through a 
thermodynamically motivated gradient dynamics formulation, Eq.~(\ref{eq:GDTF}). 
The elastic energy contribution acts in a stabilising manner in each of these models, 
consistent with the physically-motivated arguments of, e.g., 
Ref.~\onlinecite{Mechkov2009a, Vandenbrouck1999}.

In contrast, the long-wave models of 
Refs.~\onlinecite{BenAmar2001, Cummings2004, Cummings2011, Lin2012}, 
which use an alternative normal stress balance that is not consistent with the bulk 
equations, lead to qualitatively different results. The normal stress boundary 
condition in these papers neglects the contribution of the elastic stress tensor, 
which leads to a change in sign of the elastic contribution in the free surface 
evolution equation. A third approach used by Carou et al.~\cite{Carou2005, 
Carou2006, Carou2007} scales the nematic elasticity such that, to leading order, 
the free surface is unaffected by the elasticity, and one recovers the Newtonian 
thin film equation, Eq.~(\ref{eq:TFCDMW}).

One should note that the strong anchoring models presented here are only valid 
for rather thick films as noted in Sec.~\ref{sec:INTRO}. First, the main assumption 
of the model -- the strong anchoring of the director at both interfaces -- is only 
valid for $h\gg h_c$ where $h_c$ is defined in Eq.~(\ref{eq:critical}). For moderate 
film thickness $h\approx h_c$ or even thinner, the surface anchoring energy has 
to be taken into account (See Ref.~\onlinecite{Barratt1973} 
and~\onlinecite{Jenkins1974} for the interfacial boundary conditions of static NLC). 
This may be done via an ad-hoc amendment of the free 
surface anchoring condition in Eq.~(\ref{eq:2DS2B1}) (cf. the approach taken 
in Ref.~\onlinecite{Cummings2011}, for the model with the `Newtonian' normal 
stress balance). Alternatively it may be modelled via the variational approach which 
will be the subject of future work. 

Second, as for isotropic liquids with film thickness below about 100 nm, long- and 
short-range effective intermolecular forces between the substrate and the free 
surface have to be taken into account possibly through a Derjaguin or disjoining 
pressure that describes wettability effects~\cite{Starov2009}. For nematic liquid 
crystals the influence of van der Waals interactions has been discussed, e.g., in 
Ref.~\onlinecite{Effenterre2003, Ziherl2003b}. Additional Casimir-type forces may 
be induced by fluctuations of the director orientation, most notably in very thin 
films with uniform director orientation, i.e., in the planar (P) state~\cite{Ziherl2000, 
Ziherl2003b}. Note, however, that the notions ``disjoining pressure'' or ``structural 
disjoining pressure'' are used in Refs.~\onlinecite{Perez1977, Rey2001, Rey2007} 
to denote the pressure contribution resulting from the elasticity of the liquid crystal, 
i.e., the last term in Eq.~(\ref{eq:GDTF}).

We would also like to point out that, within the present long-wave scalings 
(Eq.~(\ref{eq:2DSCAL})), there is no distinction in the elastic energy whether 
the director is bent clockwise or counterclockwise. The two director profiles 
shown in Fig.~\ref{FIG:director}(b) (on the right hand side and on the left hand 
side of the dashed (black) line) have exactly the same elastic energy, 
$\widetilde{K}/2h^2$. However, such a situation is still not allowed even 
though the elastic energy is continuous across the dashed (black) line. The 
director field is discontinuous and it breaks the long-wave assumption 
($\partial^2_x\theta \ll \partial^2_z\theta$). A simple way to circumvent this was 
proposed in Ref.~\onlinecite{Lin2012}, whereby the discontinuity of 
Fig.~\ref{FIG:director}(b) is smoothed out over a given range. More sophisticated 
models for real defects are needed. For instance, one may incorporate a 
description of the dynamics of the scalar order parameter related to the 
nematic-isotropic transition. Away from the phase transition it can be employed to 
model defects. Such a model would also allow one to tackle the structuring of 
films that occurs close to the nematic-isotropic 
transition~\cite{Herminghaus1998,Schlagowski2002,Effenterre2003,Garcia2008}.

\begin{figure}[tbh]
  \begin{center}
    \subfigure[Coordinates]{\includegraphics[height=5cm]{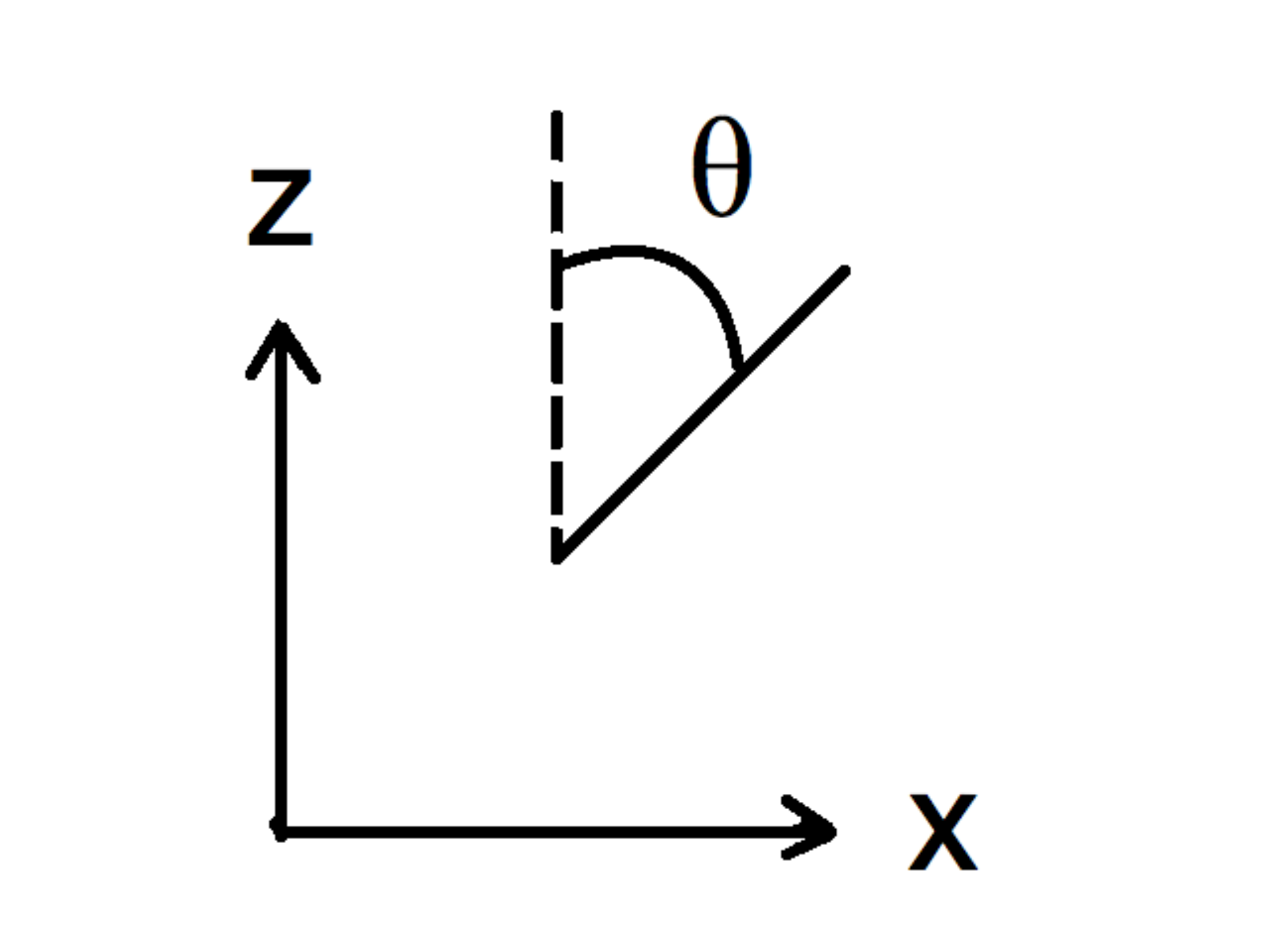}}
    \subfigure[Director orientation]{\includegraphics[height=5cm]{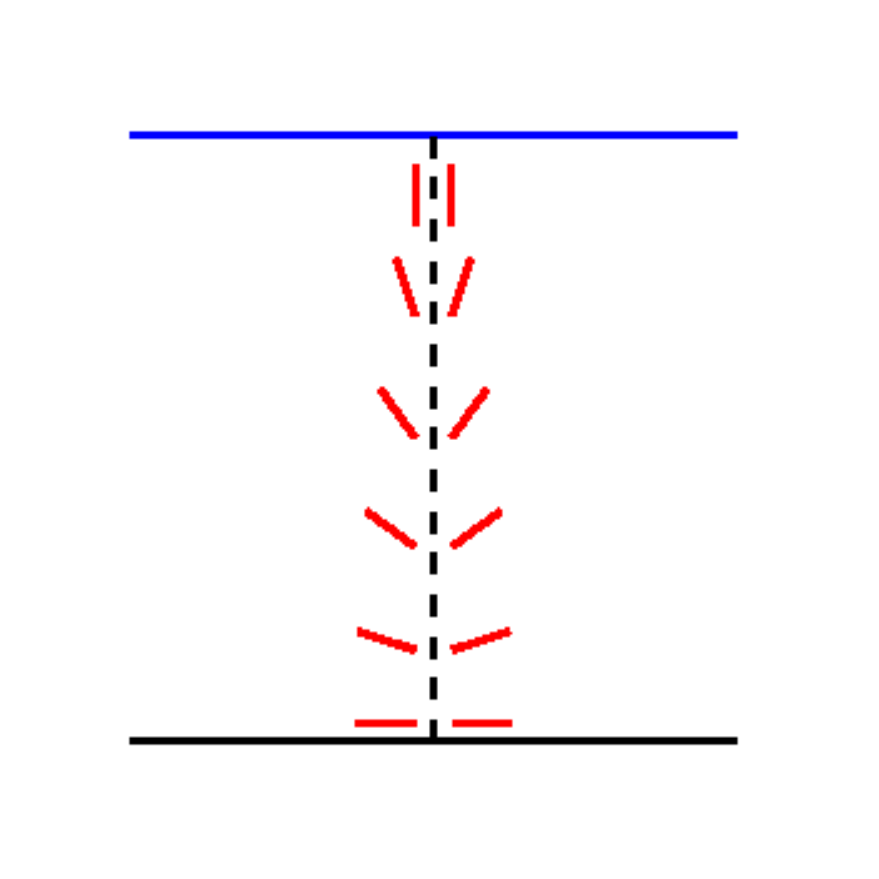}}
  \end{center}
  \caption{(Color online) The coordinates used in this manuscript are given in (a). 
  The angle $\theta$ of the director is measured with respect to the positive $z$ 
  axis. In (b), we present two possible director profiles of a hybrid film. The 
  molecules can bend either clockwise or counterclockwise across the film. The 
  solid (blue) top curve indicates the free surface, the solid (black) bottom curve 
  indicates the solid substrate, short (red) lines represent the orientation of the 
  director field, and the dashed (black) line indicates the defect location.}
  \mylab{FIG:director}
\end{figure}

In conclusion, we have clarified how the elastic contribution influences the free 
surface of a nematic film under the strong anchoring assumptions. Within the 
long-wave scalings, we have discussed two cases, corresponding to weak and 
moderate elasticity, respectively:
\begin{itemize}
\item $\frac{K}{\mu U L}=O\left(\delta\right)$: The bulk elasticity has only a minor 
influence on the free surface evolution. It does not affect the stability of a film. The 
evolution of the film and the director field are given by Eq.~(\ref{eq:S1TF1}) and 
Eq.~(\ref{eq:S1TF3}), respectively. 

\item $\frac{K}{\mu U L}=O\left(1\right)$: The strong antagonistic anchoring makes 
a significant contribution leading to a diffusion-like term in the film surface height 
evolution equation - see Eq.~(\ref{eq:GDTF}). Furthermore, the director always 
maintains its steady state, given by Eq.~(\ref{eq:director}). 
\end{itemize}
The models can be derived either from asymptotic expansion of the 
nemato-hydrodynamic equations or from a thermodynamically motivated gradient 
dynamics formulation. 
The former approach has the advantage of mathematical rigorousness, while 
the latter approach is much simpler in deriving the evolution equations. 
It is found that the elastic distortion energy is always stabilising.

\begin{acknowledgements}
  TL and UT acknowledge support from the European Union under Grant 
  No.\ MRTN-CT-2004-005728 (MULTIFLOW). UT would like to thank G\"unter 
  Gr\"un for an invitation to Bonn in October 2004 where they did a 
  derivation similar to that in the appendix and discussed the sign of the elasticity 
  term in the resulting evolution equation. LJC and LK acknowledge support from 
  the NSF under awards DMS-0908158 and DMS-1211713, and LJC also 
  acknowledges support from KAUST under Award No. KUK-C1-013-04.
\end{acknowledgements}

\appendix

%%%%%%%%%%%%%%%%%%%%%%%%%%%%%%%%%%%%%%%%%
\section{Long wavelength approximation of a thin film of NLC}
\mylab{app:longwave}
%%%%%%%%%%%%%%%%%%%%%%%%%%%%%%%%%%%%%%%%%

\subsection{Ericksen-Leslie equations in two spatial dimensions}

Assume the flow is two dimensional and $y$-independent, then the director field can 
be expressed as $\vect{n}=(\sin\theta,\cos\theta)^T$ and the velocity field is 
$\vect{v}=(u,w)^T$. The elastic energy reduces to 
\begin{equation}
w_F = \frac{K}{2}\left((\partial_x\theta)^2+(\partial_z\theta)^2\right).
\end{equation}
Without fluid inertia, the linear momentum equations are then given by 
(from Eq.~(\ref{eq:ELKL}))
\begin{eqnarray}
 \partial_x\left[ p + \frac{K}{2}\left((\partial_x\theta)^2+(\partial_z\theta)^2\right)\right] 
 &=& 
-K\,\partial_x\theta\,\left(\partial^2_x\theta+\partial^2_z\theta\right) 
+\partial_x\sigma^V_{11} +\partial_z\sigma^V_{12}, 
\mylab{eq:ELE1} \\
 \partial_z\left[ p + \frac{K}{2}\left((\partial_x\theta)^2+(\partial_z\theta)^2\right)\right] 
 &=& 
-K\,\partial_z\theta\,\left(\partial^2_x\theta+\partial^2_z\theta\right) 
+\partial_x\sigma^V_{21} +\partial_z\sigma^V_{22}.
\mylab{eq:ELE2}
\end{eqnarray}
(The viscous stress tensor, $\vect{\sigma}^V$, is defined in Appendix~\ref{sec:VIS2D}, 
Eq.~(\ref{eq:VST2D}).) For the angular momentum equation, 
Eq.~(\ref{eq:ELKA}), one can eliminate the Lagrange multiplier $\lambda$ by 
performing an inner product with the vector 
$\vect{n}^{\bot}=(\cos\theta, -\sin\theta)^T$. We then have
\begin{eqnarray}
 K(\partial^2_x\theta+\partial^2_z\theta) &=& 
 \gamma_1\left[\dot{\theta}-\frac{1}{2}(\partial_zu-\partial_xw)\right] \nonumber\\
& &+\frac{\gamma_2}{2}\left[(\partial_xu-\partial_zw)\sin(2\theta) 
+ (\partial_zu+\partial_xw)\cos(2\theta)\right].
\mylab{eq:ELE3}
\end{eqnarray}
The continuity equation, Eq.~(\ref{eq:ELKI}), is rewritten as
\begin{equation}
 \partial_xu + \partial_zw = 0. \mylab{eq:ELE4}
\end{equation}

\subsubsection{Boundary conditions}

In 2D, the boundary conditions for the director field, assuming strong anchoring at 
both interfaces (planar at the substrate and homeotropic at the free surface), are
\begin{equation}
\theta(z=0)=\frac{\pi}{2}, 
\quad \theta(z=h)=\cos^{-1}\left(\frac{1}{\sqrt{1+(\partial_xh)^2}}\right).
\end{equation}
For the velocity field, we assume no-slip at the solid substrate,
\begin{equation}
u=w=0, \quad \mbox{at} \quad z=0.
\end{equation}
At the free surface we have the kinematic boundary condition
\begin{equation}
w = \partial_th + u\partial_xh,  \quad \mbox{at} \quad z=h,
\end{equation}
which can be combined with the incompressibility condition, Eq.~(\ref{eq:ELE4}), 
to be 
\begin{equation}
\partial_th+\partial_x \left(\int^h_0 u\,dz\right) =0, 
\end{equation}
or equivalently, 
\begin{equation}
\partial_th+\partial_x \left(\int^h_0 \partial_zu\, (h-z) \,dz\right) =0.
\mylab{eq:THmass}
\end{equation}
(Note that the no-slip boundary condition, $u(z=0)=0$ was imposed in deriving 
Eq.~(\ref{eq:THmass}).) 

For the balance of normal and tangential stresses, we first note that the stress 
tensor for a NLC film is written as
\begin{equation}
\vect{\sigma} = -p
\left[\begin{array}{cc}
1 & 0\\
0 & 1
\end{array}\right]
-K\left[\begin{array}{cc}
(\partial_x\theta)^2 & \partial_x\theta\,\partial_z\theta \\
\partial_x\theta\,\partial_z\theta & (\partial_z\theta)^2
\end{array}\right]
+
\left[\begin{array}{cc}
\sigma^V_{11} & \sigma^V_{12} \\
\sigma^V_{21} & \sigma^V_{22}
\end{array}\right]
\end{equation}
and the stress tensor of the air phase is $\vect{\sigma}^i = -p_0\vect{I}$. We 
assume that the jump in the normal stress is balanced by surface tension and the 
jump in tangential stress is zero. That is
\begin{equation}
\vect{k}\cdot (\vect{\sigma} - \vect{\sigma}^i) \cdot \vect{k} = \gamma \kappa, \quad 
\vect{k}\cdot(\vect{\sigma} - \vect{\sigma}^i) \cdot \vect{t} = 0,
\end{equation}
where $\kappa$ is the curvature, $\vect{k}$ is 
the normal vector at the free surface and $\vect{t}$ is the tangent vector at the 
free surface, defined as 
\begin{equation}
\kappa = \frac{\partial^2_xh}{\left(1+(\partial_xh)^2\right)^{3/2}}, \quad 
\vect{k} = \frac{1}{\sqrt{1+(\partial_xh)^2}}\,\left(-\partial_xh, 1\right)^T, \quad 
\vect{t} = \frac{1}{\sqrt{1+(\partial_xh)^2}}(1, \partial_xh)^T,
\end{equation}
respectively.

\subsection{Non-dimensionalisation and long-wave approximation}

We make the usual long-wave scalings to nondimensionalize the governing 
equations as shown in Eq.~(\ref{eq:2DSCAL}). Also we rescale the coefficients 
of nematic viscosity by the Newtonian equivalent, setting 
$\alpha_i = \mu \bar{\alpha}_i$ where $\mu=\alpha_4/2$. For the elastic constant, 
we assume $K=\epsilon \mu U L \bar{K}$ where $\epsilon$ is a parameter of order 
$o(1/\delta)$ that will be specified later. 

The leading order equations are then given by (after dropping the over-bars)
\begin{subequations}
\begin{eqnarray}
\partial_x\left(p+\frac{\epsilon K}{2}\,(\partial_z\theta)^2\right) &=&
 -\epsilon K\, \partial_x\theta\,\partial^2_z\theta 
 + \partial_z\left(q_1(\theta)\,\partial_zu\right), \mylab{eq:S1EL1}\\
\partial_z\left(p+\frac{\epsilon K}{2}\,(\partial_z\theta)^2\right) &=&
-\epsilon K\, \partial_z\theta\,\partial^2_z\theta, \mylab{eq:S1EL2}\\
\epsilon K\,\partial^2_z\theta &=& -\delta q_2(\theta) \,\partial_zu, \mylab{eq:S1EL3}\\
\partial_xu + \partial_zw &=& 0, \mylab{eq:S1EL4}
\end{eqnarray}
\end{subequations}
where $q_1(\theta)$ and $q_2(\theta)$ are related to the viscous stress tensor, 
their full expressions are given later in Sec.~\ref{sec:VIS2D}. The leading order 
boundary conditions are the kinematic boundary condition, Eq.~(\ref{eq:THmass}), 
with 
\begin{equation}
\theta(z=0)=\frac{\pi}{2}, \quad \theta(z=h)=0, 
\mylab{eq:S1BC1}
\end{equation}
\begin{equation}
p = p_0-C\,\partial^2_xh-\epsilon K\, (\partial_z\theta)^2, \quad 
-\epsilon K (\partial_x\theta\,\partial_z\theta + (\partial_z\theta)^2 \,\partial_xh) 
+ q_1(\theta)\,\partial_zu=0, 
\quad \mbox{at $z=h$},
\end{equation}
where $C=\delta^3 \gamma/\mu U$ is the inverse capillary number.

\subsubsection{Weak elasticity ($\epsilon=\delta$)} \mylab{sec:NLC2DS1}

Assuming the elastic free energy is weak compared to the pressure, we can 
choose $\epsilon=\delta$. Observing that Eq.~(\ref{eq:S1EL2}) reduces to 
$p_z=0$ at leading order, one can solve the pressure exactly and the 
velocity is then determined by
\begin{equation}
\partial_zu(x,z) = \frac{C}{q_1(\theta)}\, (h-z)\, \partial^3_xh.
\end{equation}
Hence, by using Eq.~(\ref{eq:THmass}), we obtain the film evolution equation as
\begin{equation}
h_t + C\,\partial_x\left( Q(h) \,\partial^3_xh\right) =0,
\mylab{eq:S1TF1}
\end{equation}
where
\begin{equation}
Q(h) = \int^h_0 \frac{(h-z)^2}{q_1(\theta)} \,dz.
\mylab{eq:S1TF2}
\end{equation}
In addition, the director field satisfies
\begin{equation}
\partial^2_z\theta = \frac{C}{K}\, \frac{q_2(\theta)}{q_1(\theta)}\,(z-h)\,\partial^3_x h
\mylab{eq:S1TF3}
\end{equation}
with boundary conditions defined in Eq.~(\ref{eq:S1BC1}). 

One can see that the nematic elasticity as well as viscosity only have influence on 
the mobility function $Q$, and thus have no influence on the stability of a free 
surface. This formulation has been studied extensively by Carou 
{\it et al.}~\cite{Carou2005, Carou2006, Carou2007} both analytically and 
numerically under the assumption of small director variation.

\subsubsection{Moderate elasticity ($\epsilon=1$)} 
\mylab{sec:NLC2DS2}

On the other hand, if we have $\epsilon=1$, Eq.~(\ref{eq:S1EL3}) reduces to 
$\partial^2_z\theta=0$ at leading order and hence the director field reaches a linear 
profile in $z$ as shown in Eq.~(\ref{eq:director}). Moreover, Eqs.~(\ref{eq:S1EL1}) 
and 
(\ref{eq:S1EL2}) are simplified to 
\begin{subequations}
\begin{eqnarray}
 \partial_x\left(p+\frac{\widetilde{K}}{2h^2}\right) &=& 
 \partial_z\left(q_1(\theta)\,\partial_zu\right), \\
 \partial_zp &=& 0,
\end{eqnarray}
\end{subequations}
with boundary conditions at the free surface
\begin{equation}
p = p_0-C\,\partial^2_xh-\frac{\widetilde{K}}{h^2}, \quad \partial_zu=0.
\end{equation}
We can therefore solve the pressure and velocity field as
\begin{equation}
p(x,z) = p_0-C \,\partial^2_xh-\frac{\widetilde{K}}{h^2}, \quad 
\partial_zu(x,z) = \left(-C\,\partial^3_xh+\frac{\widetilde{K}}{h^3}\,\partial_xh\right)
\left(\frac{z-h}{q_1(\theta)}\right).
\end{equation}
As a result, the film evolution equation is given by 
\begin{equation}
h_t + \partial_x\left[Q(h)\, \left(C\,\partial^3_xh
-\frac{\widetilde{K}}{h^3}\,\partial_xh\right)\right] =0,
\mylab{eq:TFNLC}
\end{equation}
where $Q(h)$ can be evaluated explicitly as~\cite{BenAmar2001}
\begin{equation}
Q(h) = Q_0\,h^3, \quad Q_0 = \left(\frac{2}{\pi}\right)^3\int^{\pi/2}_0 
\frac{\xi^2}{q_1\left(\xi\right)} \,d\xi.
\mylab{eq:S2TF2}
\end{equation}

%%%%%%%%%%%%%%%%%%%%%%%%%%%%%%%%%%%%%%%%%
\section{Viscous stress tensor of nematic liquid crystal in two dimension} 
\mylab{sec:VIS2D}
%%%%%%%%%%%%%%%%%%%%%%%%%%%%%%%%%%%%%%%%%

The viscous stress tensor of NLC in 2D is written as
\begin{eqnarray}
\vect{\sigma}^V &=& 
\alpha_1 
\left(\sin^2\theta\,\partial_xu + \cos^2\theta\,\partial_zw 
+\sin\theta\cos\theta(\partial_zu+\partial_xw)\right) 
\left[\begin{array}{cc}
\sin^2\theta &  \sin\theta\cos\theta \\
\sin\theta\cos\theta & \cos^2\theta
\end{array}\right] 
\mylab{eq:VST2D}\\
& & +\alpha_2 \left(\frac{D}{Dt}\theta-\frac{\partial_zu-\partial_xw}{2}\right)
\left[\begin{array}{cc}
\sin\theta\cos\theta & \cos^2\theta \\
-\sin^2\theta & -\sin\theta\cos\theta
\end{array}\right] \nonumber\\
& & +\alpha_3 \left(\frac{D}{Dt}\theta-\frac{\partial_zu-\partial_xw}{2}\right)
\left[\begin{array}{cc}
\sin\theta\cos\theta & -\sin^2\theta \\
-\cos^2\theta & -\sin\theta\cos\theta
\end{array}\right] \nonumber\\
& & +\frac{\alpha_4}{2}
\left[\begin{array}{cc}
2\partial_xu & \partial_zu+\partial_xw \\
\partial_zu+\partial_xw & 2\partial_zw
\end{array}\right] \nonumber\\
& & +\frac{\alpha_5}{2}
\left[\begin{array}{cc}
2\sin^2\theta \,\partial_xu+\sin\theta\cos\theta (\partial_zu+\partial_xw) & 
2\sin\theta\cos\theta \,\partial_xu + \cos^2\theta (\partial_zu+\partial_xw) \\
\sin^2\theta (\partial_zu+\partial_xw) + 2\sin\theta\cos\theta \,\partial_zw & 
\sin\theta\cos\theta (\partial_zu+\partial_xw) + 2\cos^2\theta \,\partial_zw
\end{array}\right] \nonumber\\
& & +\frac{\alpha_6}{2}
\left[\begin{array}{cc}
2\sin^2\theta \,\partial_xu+\sin\theta\cos\theta (\partial_zu+\partial_xw) & 
\sin^2\theta (\partial_zu+\partial_xw) + 2\sin\theta\cos\theta \,\partial_zw \\
\cos^2\theta (\partial_zu+\partial_xw) + 2\sin\theta\cos\theta \,\partial_xu & 
\sin\theta\cos\theta (\partial_zu+\partial_xw) + 2\cos^2\theta \,\partial_zw
\end{array}\right]. \nonumber
\end{eqnarray}
Similarly, the coupling term, $\vect{g}$, between the director and velocity field can 
be written as
\begin{equation}
\vect{g} = -\gamma_1 \left(\dot{\theta}-\frac{\partial_zu-\partial_xw}{2}\right)
\left[\begin{array}{c}
\cos\theta \\ 
-\sin\theta
\end{array}\right]
-\frac{\gamma_2}{2}
\left[\begin{array}{c}
2\sin\theta \,\partial_xu + (\partial_zu+\partial_xw)\cos\theta \\
(\partial_zu+\partial_xw)\sin\theta+2\cos\theta \,\partial_zw
\end{array}\right]. \mylab{eq:CT2D}
\end{equation}
We also note that, within the long-wave scalings [Eq.~(\ref{eq:2DSCAL})], to 
leading order we have
\begin{equation}
\sigma^V_{12} = \mu q_1(\theta) \,\partial_zu + O\left(\frac{\mu U}{L}\right), \quad 
\vect{n}^{\bot}\cdot\vect{g} = -\mu q_2(\theta) \,\partial_zu + O\left(\frac{\mu U}{L}\right),
\end{equation}
where $\mu=\alpha_4/2$, $\vect{n}^{\bot}=(\cos\theta, -\sin\theta)^T$ and 
\begin{eqnarray}
\mu q_1(\theta) &=& \frac{1}{2}\left[\alpha_4+2\alpha_1\sin^2\theta\cos^2\theta
+(\alpha_5-\alpha_2)\cos^2\theta+(\alpha_3+\alpha_6)\sin^2\theta\right], \\
\mu q_2(\theta) &=& \frac{1}{2}\left[\gamma_1-\gamma_2\cos(2\theta)\right].
\end{eqnarray}
As an example, for Newtonian fluids, $q_1(\theta)=1$ and $q_2(\theta)=0$.

%%%%%%%%%%%%%%%%%%%%%%%%%%%%%%%%%%%%%%%%%
\section{Notation conventions} 
\mylab{sec:NOTA}
%%%%%%%%%%%%%%%%%%%%%%%%%%%%%%%%%%%%%%%%%
For clarity, we list all the notations used. We write for a vector $\vect{n}=n_i$ or 
$\vect{m}=m_i$, for a tensor $\vect{\sigma}=\sigma_{ij}$ or 
$\vect{\kappa}=\kappa_{ij}$; as the superscript $T$ denotes transposition one has 
$\vect{\sigma}^T=\sigma_{ji}$. Further, $\epsilon_{ijk}$ is the alternator. The 
notations for operators and products are $\nabla\vect{n} = n_{j,i}$, 
$\nabla\cdot\vect{n} = n_{k,k}$, $\nabla\times\vect{n} = \epsilon_{ilk}n_{k,l}$, 
$\Delta\vect{n} = n_{i,kk}$, 
$\nabla\cdot\vect{\sigma} = \sigma_{ik,k}$, 
$\vect{\sigma}\cdot\vect{\kappa} = \sigma_{ik}\kappa_{kj}$, 
$\vect{\sigma}:\vect{\kappa} = \sigma_{kl}\kappa_{lk}$, 
$\vect{\sigma}\cdot\vect{n} = \sigma_{ik} n_{k}$, 
$\vect{n}\times\vect{m} = \epsilon_{ilk}n_k m_l$, 
where `$,i$' denotes the partial derivative with respect to the $i$th component.

\bibliography{NLC_SA} 

\end{document}